\documentclass[
 twocolumn,
 amsmath,amssymb,
 aps, prb,
]{revtex4-2}

\usepackage{graphicx}
\usepackage{dcolumn}
\usepackage{bm}
\usepackage{float}
\usepackage{caption}
\usepackage{subcaption}
\usepackage{ragged2e}
\usepackage{braket}
\usepackage{xcolor}  
\usepackage[colorlinks=true, citecolor=blue, linkcolor=blue, urlcolor=blue]{hyperref}

\usepackage{ulem}

\usepackage{ulem}
\begin{document}

\preprint{APS/123-QED}

\title{Substitution modulated transition from semimetal to superconductor in ZrTe$_{2-x}$Se$_x$ with coexistence of nontrivial electronic topology}

\author{V. M. Fim}
\email{vitor.m.fim@gmail.com}
\affiliation{Escola de Engenharia de Lorena - DEMAR, Universidade de São Paulo, 12612-550, Lorena, Brazil}
\author{C. F. Schuch}
\email{caua.schuch@usp.br}
\affiliation{Escola de Engenharia de Lorena - DEMAR, Universidade de São Paulo, 12612-550, Lorena, Brazil}

\author{L. R. de Faria}
\affiliation{Escola de Engenharia de Lorena - DEMAR, Universidade de São Paulo, 12612-550, Lorena, Brazil}

\author{F. A. Santos}
\affiliation{Universidade Federal do ABC, Centro de Ciências Naturais e Humanas, Santo André, SP, Brazil}

\author{M. S. da Luz}
\affiliation{Instituto de Ciências Tecnológicas e Exatas, Universidade Federal do Triângulo Mineiro, Uberaba, Minas Gerais, Brazil}

\author{S. S. Tsirkin}
\affiliation{Chair of Computational Condensed Matter Physics, Institute of Physics,
École Polytechnique Fédérale de Lausanne (EPFL), CH-1015 Lausanne, Switzerland}

\author{L. T. F. Eleno}
\affiliation{Escola de Engenharia de Lorena - DEMAR, Universidade de São Paulo, 12612-550, Lorena, Brazil}

\author{A. J. S. Machado}
\email{ajefferson@usp.br}
\affiliation{Escola de Engenharia de Lorena - DEMAR, Universidade de São Paulo, 12612-550, Lorena, Brazil}

\date{\today}

\begin{abstract}
This study explores the emergence of superconductivity in high-quality ZrTe$_{2-x}$Se$_x$ crystals, grown via the isothermal chemical vapor transport (ICVT) technique. Resistive, structural, and thermal measurements reveal that substituting Te with Se in the ZrTe$_2$ matrix induces a superconducting state at low temperatures. The critical temperature ($T_c$) exhibits a clear dependence on the selenium concentration, peaking at $x=0.15$ with a $T_c$ of $4.8$ K. Calorimetric data indicates that even a low Se substitution range is capable of modifying both the electronic contribution and the vibrational modes of the crystal lattice. Combined with ab initio calculations and Wannier Hamiltonian interpolation between ZrTe$_2$/ZrSe$_2$, we established an extensive phase diagram mapping the transition from charge density wave (CDW) to the state with coexistence between the Dirac semimetal and superconductivity (SC), up to the semiconductor phase. This coexistence suggests that ZrTe$_{1.85}$Se$_{0.15}$ could be a candidate platform for topological superconductivity, as it hosts a nontrivial $\mathbb Z_2$ invariant, with nonvanishing surface states in its $(001)$ planes.
\end{abstract}

\maketitle


\section{Introduction}
Transition metal dichalcogenides (TMDs), compounds with the general formula $MX_2$, where $M$ is a transition metal (e.g., Zr, Hf, Ti, Mo, W, Ta) and $X$ is a chalcogen atom (S, Se, or Te), represent an intriguing class of low-dimensional (2D) materials. These materials are characterized by their layered crystal structures, where metal layers are sandwiched between two chalcogen layers and held together by weak van der Waals forces \cite{1}. This intrinsic bidimensionality, combined with accessible interlayer sites, enables intercalation of various atoms and molecules and provides TMDs with exceptional tunability for a wide range of applications, from electrocatalysis and batteries to spintronics and other quantum technologies \cite{1,18,19,20,Montblanch2023LayeredMaterials}.

Within these systems, the emergence and complex interplay of competing electronic instabilities such as charge density waves (CDW) and superconductivity (SC) are frequently observed \cite{2,3e22,4,5}. The delicate balance between these collective phenomena often gives rise to rich phase diagrams, exemplified by the coexistence of CDW/SC in Ni-doped ZrTe$_2$ \cite{3e22}, exciton condesation in 1T-ZrTe$_2$ \cite{Gao2023ExcitonCondensation}, or by the intricate interplay of superconductivity and magnetism in monolayers \cite{PhysRevB.95.104515}, also exotic magnetic spin states such as the spin-glass phase of Fe$_x$NbSe$_2$ \cite{c87z-wrhj}. Beyond these electronic instabilities, many $MX_2$ compounds also exhibit nontrivial topology in their electronic band structure, characterized by Dirac cones in reciprocal space \cite{6,PhysRevB.101.165122}. The universal presence of such bulk Dirac cones and topological surface states, emerging from single orbital manifolds, imparts unique electronic properties to these materials, making them highly attractive for next-generation quantum electronics and spintronics applications \cite{6}.

While stoichiometric ZrTe$_2$ is a semimetal and does not exhibit superconductivity down to $1.8\,\mathrm{K}$, slight deviations from stoichiometry can profoundly alter its electronic ground state. For instance, tellurium-deficient ZrTe$_{1.8}$ unexpectedly displays superconductivity with a critical temperature $T_c = 3.2\,\mathrm{K}$ \cite{7}. This intriguing behavior suggests a multiband superconducting mechanism, with superconductivity emerging predominantly within the interlayer (van der Waals gap) regions of the material.

\begin{figure}
    \includegraphics[width=1.\linewidth]{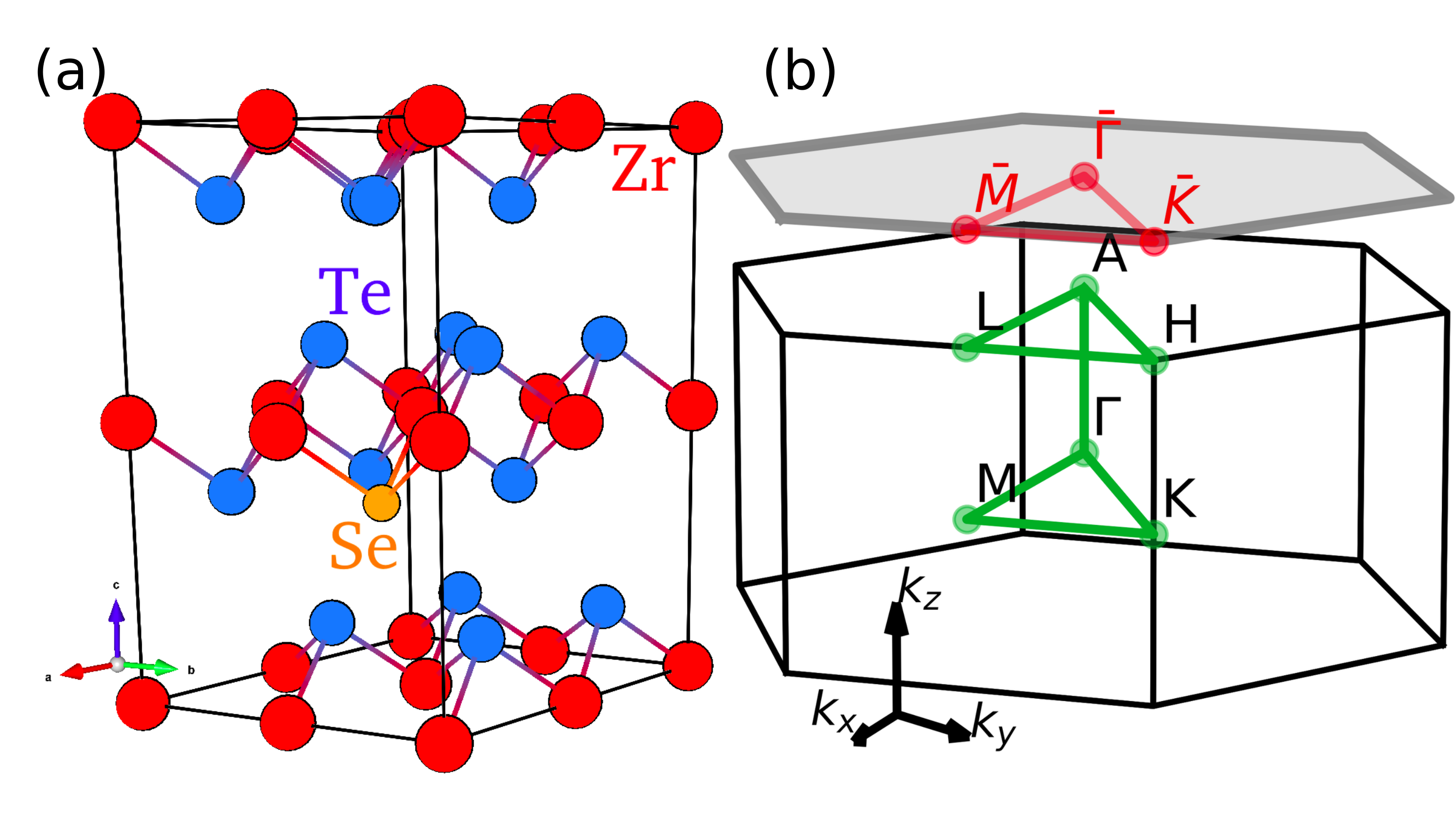}

    \caption{\justifying(a) Crystal structure of ZrTe$_{2-x}$Se$_{x}$ in the $2\times2\times2$ supercell employed in DFT calculations, representing the disordered composition $x=0.125$; (b) 1st Brillouin Zone of SG 164 with highlighted path (green) used to plot the dispersion curves, and the projected points $\bar K\bar M\bar\Gamma\bar K$ (red) in its $(001)$ plane.}
    \label{fig:sg164data}
\end{figure}

Motivated by this compositional tunability, the present work focuses on the ZrTe$_{2-x}$Se$_x$ system, depicted in Fig. \ref{fig:sg164data}(a), investigating the effects of substituting Te with Se within the ZrTe$_2$ matrix, which crystallizes in the CdI$_2$-type $P\bar{3}m1$ (No.~164) structure \cite{8}. Importantly, despite selenium atoms being approximately $19\%$ smaller by linear size than tellurium atoms, this substitution within the van der Waals gaps is structurally accommodated. This structural compatibility is further supported by previous observations in the related layered system ZrTe$_{3-x}$Se$_x$, where a similar substitution also induces superconductivity with $T_c = 4.5\,\mathrm{K}$ \cite{9}. Increasing Se content in ZrTe$_2$ is known to drive a transition from semimetallic to semiconducting behavior for higher substitution levels ($x > 1$) \cite{10}, further highlighting the broad electronic tunability of this system. Additionally, the strong relationship between structural modifications, either via intercalation or substitution, and the emergence of superconductivity in ZrTe$_2$- and ZrTe$_3$-based materials has been widely documented \cite{11,13}. Despite extensive investigations of the ZrTe$_2$–ZrSe$_2$ transition, no superconductivity has been reported so far.

With this study, we use both theory/experiment to extend previous works \cite{PhysRevB.101.165122,10} and systematically explore the impact of controlled Se substitution for Te on the electronic properties of ZrTe$_2$. Our objective is to demonstrate that superconductivity can be effectively induced and tuned through these subtle structural and electronic modifications. High-quality single crystals were synthesized using the isothermal chemical vapor transport (ICVT) technique, ensuring precise control over selenium incorporation. The emergence of superconductivity was thoroughly characterized using electrical resistivity and calorimetric measurements. Our findings reveal the existence of a superconducting dome in ZrTe$_{2-x}$Se$_x$, allowing us to construct a comprehensive phase diagram that clearly delineates the charge density wave, semimetal and superconducting regions as a function of selenium concentration. Employing first principle calculations, we describe the electronic structure evolution within supercell calculations and Wannier-based Hamiltonians, in which we interpolate between the pristine real space Hamiltonians of ZrTe$_2$ and ZrSe$_2$ to provide a deeper understanding of the nontrivial persisting topological phase and semimetal to semiconductor transition. First, we found the coexistence of Dirac semimetallic and superconducting states, up to a value of $x=0.3$, and, using slab Bogoliubov-de-Gennes Hamiltonian, studied its ground state and explore the nontrivial topological phases via heterostructure engineering.

\section{Experimental methods}
ZrTe$_{2-x}$Se$_x$ single crystals were synthesized using the isothermal chemical vapor transport (ICVT) technique \cite{14}, known for producing high-quality and phase-pure crystals \cite{2,3e22,4,6}. For this purpose, polycrystalline pellets with nominal composition ZrTe$_{2-x}$Se$_x$ were pre-synthesized by solid-state reaction and sealed in a quartz ampoule under vacuum with approximately $2.5\,\mathrm{mg/cm^3}$ of iodine, which served as the transport agent. The ampoule was placed horizontally in a box furnace and maintained at a temperature of $1050\,^{\circ}\mathrm{C}$ for 7 days. As a result, crystals grew directly from the pellet, with the largest specimens typically reaching dimensions of $10 \times 10 \times 0.1\,\mathrm{mm^3}$.

The chemical composition of the synthesized samples was determined by energy-dispersive X-ray spectroscopy (EDS) using a Hitachi scanning electron microscope. The crystallographic quality was assessed by X-ray diffraction (XRD) using a Panalytical Empyrean diffractometer. Oscillation curves centered on the $(00l)$ reflections were used to evaluate both the crystallographic orientation and the degree of crystallinity.

Electrical resistivity and specific heat measurements were performed using a Quantum Design Physical Property Measurement System (PPMS-9 Evercool II), equipped with a $9.0\,\mathrm{T}$ superconducting magnet. Resistivity measurements were conducted using a standard four-probe configuration, with copper leads attached by silver paste along the $ab$ plane of the sample, yielding low-resistance contacts typically around $1.0\,\Omega$. Specific heat measurements were carried out using the system’s calorimetry option, which employs the thermal relaxation method.

\section{Theoretical methods}
\subsection{DFT calculations}
\label{dftmethod}

We performed the ab initio calculations using two different codes: \textsc{GPAW} \cite{GPAW1, GPAW2, GPAW3} and \textsc{Quantum ESPRESSO} (QE) \cite{giannozzi2009,Giannozzi_2017}, within the Perdew–Burke–Ernzerhof (PBE) generalized gradient approximation (GGA) parametrization of the exchange-correlation functional \cite{gga-pbe}. The Projector Augmented Wave method (PAW) \cite{PhysRevB.50.17953} was used to describe the ion-electron interactions in both GPAW (built-in pseudopotentials) and QE (pseudopotentials from \texttt{PSLIBRARY} \cite{PSlibrary}), where the same valence configuration was defined: $4s^2 4p^4$ for Se, $5s^2 5p^4$ for Te, and $4s^2 5s^2 4p^6 4d^2$ for Zr. In the first approach, we used a total of 24 structures to account for the complete transition between ZrTe$_2 \rightarrow$ ZrTe$_{2-x}$Se$_x \rightarrow$ ZrSe$_2$, where the lattice parameters for the pristine crystals ($x=0$ or $x=2$) were fixed from experimental data \cite{MAO20191,https://doi.org/10.1107/S2053229618009841}, and, for the intermediate $x$ values, the lattice parameters were linearly interpolated between them. The $2d$ Wyckoff site for the chalcogenide (Te or Se) was fixed at $(^1/_3,\, ^2/_3,\,0.25)$ for the chalcogenide $2d$ site, so no relaxation was performed, because the $z$-coordinate almost doesn't change with composition, as shown by preliminary calculations, and to allow for the Wannier interpolation (to be described in Sec. \ref{winterp}). Thus, just scalar relativistic calculations with a plane-wave cutoff energy of $500$ eV, and a $6\times6\times4$ $\Gamma$-centered $\mathbf{k}$-mesh was necessary, as \textsc{GPAW} was used solely to get the Wannier functions (WFs).  For the QE calculations, cutoff energies of $110$ and $8\times110$ Ry were applied for plane-waves and charge density of the fully-relativistic calculations, respectively. For the Density of States (DOS) determination and validation of the interpolation method, we built $2\times2\times2$ supercells using the \texttt{Supercell} program \cite{Okhotnikov2016}, which yielded $1$ and $7$ configurations for $x=0.125$ and $0.25$, respectively, but we neglected disorder effects, and chose one randomly. The supercells were then fully relaxed, using the non-local functional \texttt{vdW-DF3} \cite{vdw-d3, vdw2, vdw-review1, vdw-review2}, and a $8\times8\times8$ grid for self-consistent steps and $12\times12\times12$ for the non-self-consistent runs using the tetrahedron method \cite{PhysRevB.49.16223} for Brillouin Zone (BZ) integrals. Unfolding of the supercell was done using the \textsc{BandUp(py)} code \cite{unfold1,unfold2,IrRep}. Fermi surfaces were plotted using FermiSurfer \cite{KAWAMURA2019197}.
 
\subsection{Wannier interpolation}
\label{winterp}

The projection of Bloch wave functions onto WFs was obtained using the WannierBerri (WB) code \cite{Tsirkin2021}, with Te/Se-p and Zr-d orbitals as the initial basis, resulting in a total of $N_w=2 \times 3 + 5 = 11$ WFs. The bands were frozen in an energy window of $E_F-6\leq E \leq E_F+2.7$ eV during the disentanglement to account for the entire valence region and first conduction bands, and the spread of each WF was less than $3\,$\AA$^2$. Spin-orbit coupling (SOC) was added non self-consistently after the wannierization procedure via the SOC Hamiltonian $H_{SOC}$, to yield the fully relativistic $H_{FR}\,\space(2N_w \times 2N_w)$:
\begin{equation}
H_{FR}(\mathbf{R}) =
\begin{pmatrix}
H_{\mathrm{SR}}(\mathbf{R}) & 0 \\
0 & H_{\mathrm{SR}}(\mathbf{R})
\end{pmatrix}
+ H_{\mathrm{SOC}}(\mathbf{R}) \, ,
\end{equation}
where $H_{\mathrm{SOC}}$ is organized as:
\begin{equation}
H_{\mathrm{SOC}}(\mathbf{R}) =
\begin{pmatrix}
H_{\uparrow\uparrow}(\mathbf{R}) & H_{\uparrow\downarrow}(\mathbf{R}) \\
H_{\downarrow\uparrow}(\mathbf{R}) & H_{\downarrow\downarrow}(\mathbf{R})
\end{pmatrix} \, ,
\end{equation}
using a spin-interlaced order. This matrix was built by applying the SOC operator based only in the region inside the augmentation sphere of the PAW method, from the all-electron partial waves $\ket{\phi^a_{i\sigma}}$. So, using the GPAW SOC routine \cite{GPAW1,PhysRevB.94.235106} with the scalar-relativistic eigenstates included in the all-electron basis $\ket{\psi^a_{n\sigma}}=\sum_i\ket{\phi^a_{i\sigma}}\braket{\tilde p_{i\sigma}^a|\tilde \psi_{n\sigma}}$, we obtain the orbital part of the SOC operator:
\begin{equation}
dV^a_t=\bra{\phi_{i\sigma}^a}\frac{1}{r}\frac{dV}{dr}L_t \ket{\phi_{j\sigma}^a},
\end{equation}
and the spin-dependent contribution is added afterwards via WB combining the Pauli matrices: $H_{SOC}=\sum_tdV_t\cdot\sigma_t$. This on-site SOC approximation in the case of dominant Te (Se)-$5p\,(4p)$ and Zr-$4d$ is widely used in tight-binding models and yields excellent agreement with fully-relativistic approaches \cite{soc_val1,soc_val2}.

Furthermore, after the wannierization procedure and SOC addition, to simulate ZrTe$_{2-x}$Se$_x$ compositions we performed a linear interpolation on the Wannier Hamiltonian between the pristine compounds ZrTe$_2$ ($x=0$) and ZrTe$_2$ ($x=2$). With this method, recently implemented in \cite{Sadhukhan2026}, it is possible to obtain the Wannier Hamiltonian at any value of $x$ by linearly interpolating between 2 Wannier Hamiltonians that have the same number and character of Wannier functions.
\begin{align}
    \mathbf r_n(\alpha)=(1-\alpha)\mathbf r_n^{\text{ZrTe}_2}+\alpha\mathbf r_n^{\text{ZrSe}_2},\\X_{ij}(\mathbf R,\alpha)=(1-\alpha) X_{ij}^{\text{ZrTe}_2}(\mathbf R,\alpha)+\alpha X_{ij}^{\text{ZrSe}_2}(\mathbf R,\alpha),
\end{align}
where $\mathbf{r}_n\,,X_{ij}(\mathbf R)$ are the Wannier centers and the real-space matrices of the system (in this case, $H_{mn}(\mathbf R)=\bra{w_{n}(0)}\hat H\ket{w_{m}(\mathbf R)}$), respectively, and $\alpha=x/2$. This methodology has proven to be straightforward and computationally efficient, as only few, non-SOC DFT calculations are required. After the interpolation, Fermi energy ($E_F$) determination was made by fitting the free parameter $E$ of the cumulative density of states:
\begin{align}
N(E)=\int^E_{-\infty}n(\epsilon)d\epsilon    
\end{align}
to return the expected $N(E_F)=24$ ($12 \,e^-$ from Zr and $2\times6\,e^-$ from Te/Se valences). For reproducibility, a complete tutorial was uploaded in \cite{wannierberri-tutorial}.

Finally, we process our interpolated Wannier Hamiltonians with WannierTools \cite{WU2017} package for two purposes. First, in order to study the topologically protected surface states, we construct slab systems of 14 unit cell layers along $c$ direction, with vacuum thickness of 20 \AA. Second, we study the adiabatic charge pumping in the bulk via Wannier charge centers (Wilson loop) calculation \cite{PhysRevB.83.235401,PhysRevB.95.075146,PhysRevB.84.075119,PhysRevB.83.035108}.

\section{Results}
\subsection{Transport measurements}
\begin{figure*}[t]
    \includegraphics[width=1.\linewidth]{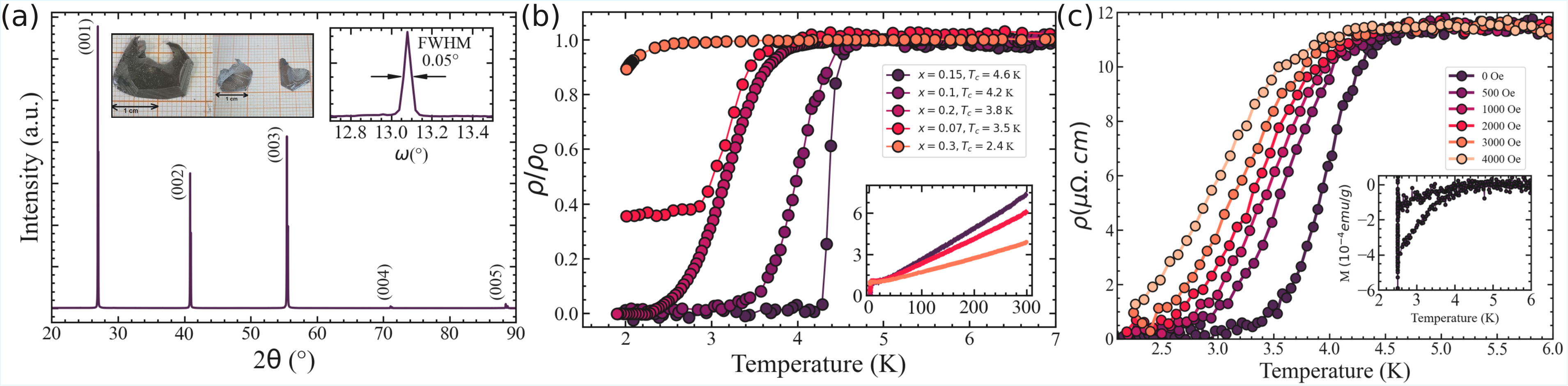}
    \caption{\justifying (a) X-ray diffraction pattern of ZrTe$_{2-x}$Se$_x$ single crystals showing only $(00l)$ reflections, consistent with preferential growth along the $c$-axis. The inset shows representative ZrTe$_{1.9}$Se$_{0.1}$ single crystals obtained by the ICVT method and the rocking curve measured around the $(003)$ reflection, yielding a full width at half maximum (FWHM) of approximately $0.05^\circ$. (b) Temperature dependence of the normalized in-plane resistivity $\rho/\rho_0$ for different Se concentrations in the low-temperature region. The inset shows the normalized resistivity between $300$ and $2$ K. (c) Resistivity as a function of temperature for ZrTe$_{1.85}$Se$_{0.15}$ measured under different applied magnetic fields. The inset shows the corresponding magnetic susceptibility measurement.}
    \label{fig:panel_expt}
\end{figure*}

The ICVT crystal growth technique proved effective in producing large, high-quality, and phase-pure ZrTe$_{2-x}$Se$_x$ crystals, with precise control over the substitution of Te by Se. EDS analysis performed on various single crystals revealed Se compositions of $x = 0.07$, $0.10$, $0.15$, $0.20$, and $0.30$. Fig. \ref{fig:panel_expt}(a) displays three representative crystals obtained after 7 days of growth, exhibiting typical dimensions of approximately $12 \times 12 \times 0.5\,\mathrm{mm^3}$.

To evaluate the crystallographic quality of the crystals, X-ray diffraction (XRD) measurements were carried out using Cu-K$\alpha$ radiation. During the measurements, the largest face of the crystal was aligned perpendicular to the incident beam. The resulting diffractogram is presented in Fig. \ref{fig:panel_expt}(a). All single crystals synthesized in this study displayed consistent diffraction patterns, featuring well-defined peaks indexed to the $(00l)$ family of planes, indicative of a preferred orientation along the $c$ axis and excellent structural quality. The absence of additional reflections further confirms the high phase purity and single-crystalline nature of the samples.

As shown in Fig. \ref{fig:panel_expt}(a), the orientation of the obtained crystals corresponds to the $(00l)$ family of planes of the CdI$_2$-type structure \cite{8}. To gain further insight into the crystal quality, an $\omega$ scan (rocking curve) was performed by decoupling the $\theta$ and $2\theta$ angles and focusing on the $(003)$ reflection peak. The observed FWHM of approximately $0.05^\circ$ indicates excellent crystallographic quality.

Figure \ref{fig:panel_expt}(b) shows the temperature-dependent normalized in-plane electrical resistivity, $\rho/\rho_0(T)$, for different Se substitution levels ($x = 0.07$, $0.10$, $0.15$, $0.20$, and $0.30$), measured in the low-temperature range from $2$ to $7\,\mathrm{K}$. Here, $\rho_0$ is defined as the normal-state resistivity at $T = 5\,\mathrm{K}$, just above the onset of the superconducting transition. All samples undergo superconducting transitions within the temperature interval $2.4$--$4.6\,\mathrm{K}$. The critical temperature $T_c$ was determined at $90\%$ of the normal-state resistivity. The high-temperature behavior, shown in the inset, reveals metallic-like behavior.
The trend of the superconducting transition temperature is not monotonic with the nominal selenium content; instead, $T_c$ increases as the partial chalcogen substitution grows from zero to $0.15$, reaches a maximum in the crystal composition ZrTe$_{1.85}$Se$_{0.15}$, and decreases upon further substitution, being completely suppressed at $0.05\%$ Se levels. The optimal composition for maximizing the superconducting transition was observed in crystals with the ZrTe$_{1.85}$Se$_{0.15}$ composition, exhibiting an onset temperature of approximately $4.8\,\mathrm{K}$. It is important to notice that Se content above $0.3$ did not exhibit any superconducting transition down to $T = 1.8\,\mathrm{K}$. In contrast, compositions near $x = 0.07$ displayed filamentary superconductivity, coexisting with charge density wave (CDW) behavior, both competing for the same electrons at the Fermi surface.

The superconducting phase diagram $T_c$ vs \%at.\ Se, built from the electrical resistance data, is shown in Figure~4, where three distinct phases can be identified, each depending on the level of Te substitution by Se in the material. It is clear that substitution of Te for Se results in the appearance of a superconducting state between the charge density wave (CDW) \cite{16} and Dirac semimetal ground states \cite{5}. The variation of the critical temperature with Se substitution reveals a superconducting dome similar to the observed when hydrostatic pressure was applied to ZrTe$_2$ \cite{14}. In that case, superconductivity appears at approximately $8.3\,\mathrm{GPa}$, with the critical temperature ($T_c$) peaking at $5.6\,\mathrm{K}$ near $19.4\,\mathrm{GPa}$, following a characteristic dome-shaped trend.

Among the superconducting chalcogenides, the pressure effect on the superconducting state is often complex. In general, the pressure leads to a suppression of the CDW state that is correlated with an enhancement, or even the emergence, of superconductivity; such a correlation highlights the competing nature of CDW and superconductivity. For instance, SC domes in 6R-TaS$_2$ and o-TaS$_3$ develop as the suppression of CDW takes place with pressure \cite{26,27}, with the suggestion of a maximum of $T_c$ occurring at a quantum critical point in o-TaS$_3$. On the other hand, the partial chalcogen substitution in Fe(Se,S) plays a key role in the emergence of a superconducting dome, probably due to a chemical pressure that is as effective as external pressure in increasing $T_c$ \cite{28}. Similar results have been reported in the Te-substituted material CuIr$_2$Te$_{4-x}$Se$_x$ \cite{29}, in 2H-TaSe$_{2-x}$S$_x$ \cite{Li2017TaSe2Sx} and 2H-NbSe$_2$ \cite{Cho2018NbSe2}. Those results highlight the two-fold aspect of the partial substitution in chalcogenides: the CDW suppression due to disorder and the intrinsic chemical pressure associated with different atomic sizes. Thus, the observed variation of $T_c$ in the ZrTe$_{2-x}$Se$_x$ crystals may come from a combined effect of disorder and chemical pressure.

Another aspect relevant to this discussion is the interplay of SC and topology, such as reported in Td-MoTe$_2$ Weyl semimetal \cite{28}. In this case, the superconducting dome arises from pressure experiments and is related to highly sensitive changes in the lattice constants of the topological phase. In the same direction, combined theoretical calculations and in situ high-pressure experiments reveal that the emergence of superconductivity coincides with a topologically nontrivial state, positioning ZrTe$_2$ as a promising platform for investigating topological superconductivity \cite{14}. The electronic behavior of ZrTe$_2$ is typical of a system which exhibits features revealed by previous scanning tunneling microscopy/spectroscopy (STM/STS) \cite{5,17}.
\begin{figure}
    \includegraphics[width=1.\linewidth]{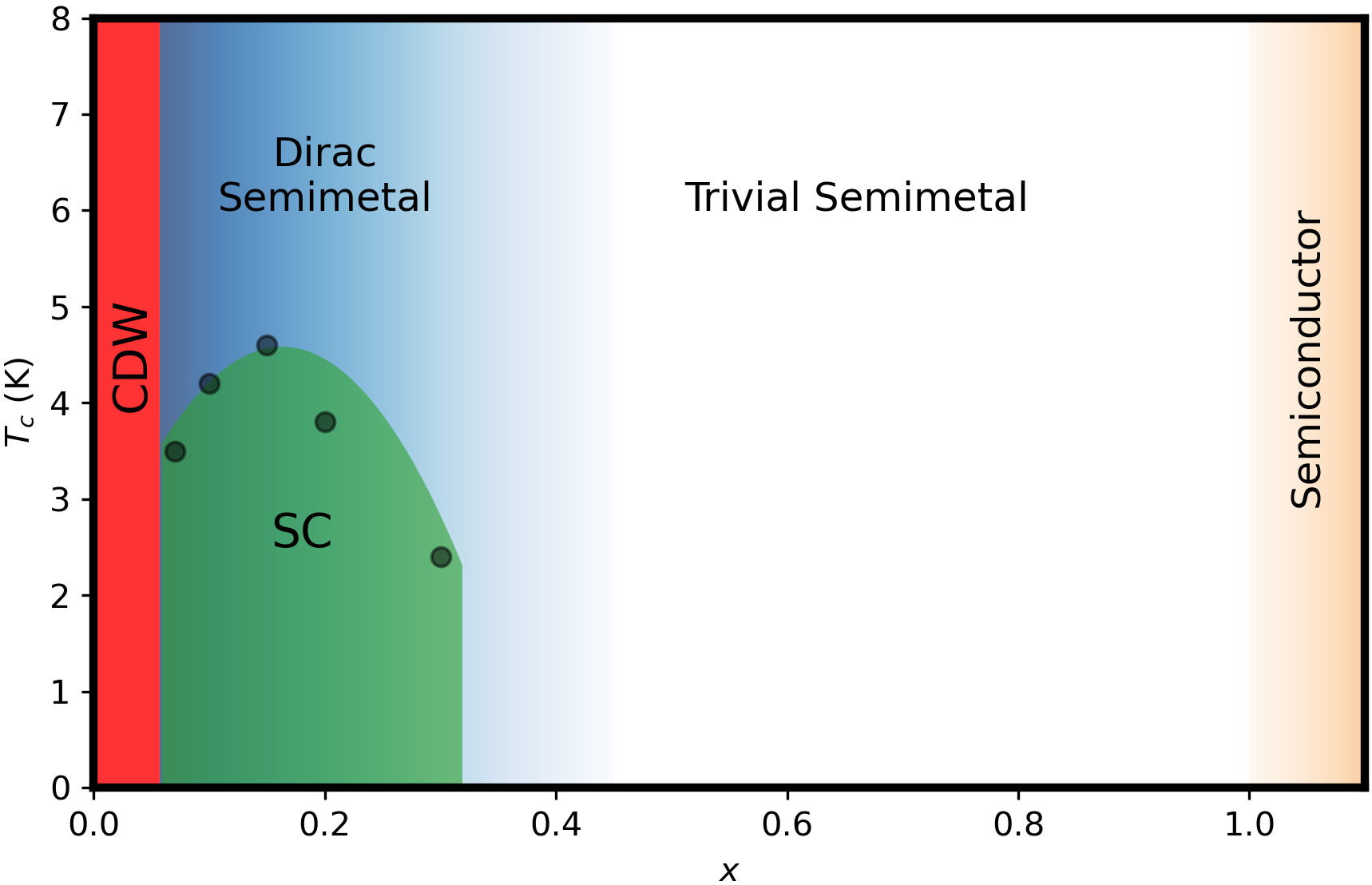}

    \caption{\justifying Proposed phase diagram of the ZrTe$_{2-x}$Se$_x$ single crystals as a function of temperature and Se content $(x)$, within the range of $x = 0$ to $x = 1$. The black points are respective to the experimental data.}
    \label{fig:phase_diagram}
\end{figure}

To gain deeper insight into the superconducting state of the ZrTe$_{2-x}$Se$_x$ system, calorimetric measurements were conducted from 300 K down to 2 K on a crystal with ZrTe$_{1.85}$Se$_{0.15}$ composition, as shown in Figure 5.
\begin{figure}
    \includegraphics[width=1.\linewidth]{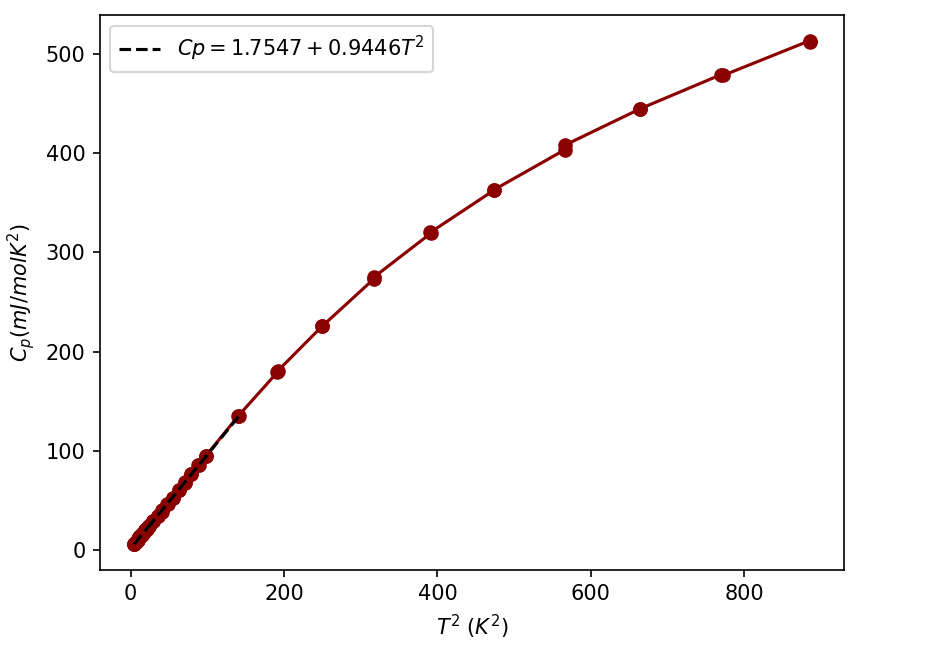}

    \caption{\justifying Measurement of $C_p/T$ ($\frac{J}{mol\,K}$) as a function of $T^2\, (K^2)$ in the range of 300-2 K, for the ZrTe$_{1.85}$Se$_{0.15}$ single crystal. The fitting of the figure is done in the region of 180-2 K, where the regime can be considered linear.}
    \label{fig:Cp_fit}
\end{figure}

Our experimental results do not support a purely filamentary superconducting scenario. As shown in Fig. \ref{fig:panel_expt}(c), the superconducting transition is gradually shifted with increasing applied magnetic field, a behavior characteristic of bulk superconductors. Furthermore, the magnified detail in the same plot reveals an weak diamagnetic response after the sample reaches the critical temperature, indicating that the observed superconductivity is not conventional in nature.

These findings suggest that the superconducting state is intrinsically topological. The absence of a clear superconducting anomaly in the specific heat ($C_p$) versus temperature ($T$) data further supports this interpretation. It is important to emphasize that this behavior is observed exclusively in ZrTe$_2$-based compounds \cite{4}, and not in analogous materials based on ZrTe$_3$ \cite{4}.

Using the Debye model in the low temperature $C_p(T)$ data, it is possible to extract the Sommerfeld constant ($\gamma$) and the harmonic phononic coefficient ($\beta$). This model can be adjusted by the observed linear behavior of $C_p/T$ vs $T^2$ (see Fig. \ref{fig:Cp_fit}) at low temperature given by equation \ref{eq:cpt}:

\begin{equation}
\frac{C_p}{T} = \gamma + \beta T^2
\label{eq:cpt}
\end{equation}

The Sommerfeld constant $\gamma$ is directly related to the $n(E_F)$ (DOS) through the relation
\begin{equation}
\gamma = \frac{2\pi^2}{3} k_B^2 n(E_F),
\label{eq:electronic_contrib}
\end{equation}
while the angular coefficient $\beta$ is related to the Debye temperature by the relation
\begin{equation}
\beta = \frac{12\pi^4 R \cdot n}{5 \Theta_D^3},
\label{eq:phonon_contrib}
\end{equation}
where $k_B$ is the Boltzmann constant, $R$ is the ideal gas constant, $\Theta_D$ represents the Debye temperature, and $n$ is the number of atoms per unit cell. The parameters extracted from the linear fit of the specific heat, in the temperature range of 0--180 K, are compiled in Table \ref{tab:exp}. The results indicate that the partial substitution of tellurium with selenium leads to a decrease in the electronic contribution ($\gamma$) compared to the pure ZrTe$_2$ compound. This finding suggests that superconductivity in the ZrTe$_{2-x}$Se$_x$ system is not driven by an increase in the density of electronic states at the Fermi level. In the conventional context, the critical temperature ($T_c$) is governed by both the electronic density of states ($\gamma$) and the electron-phonon interaction ($\beta$). The electron-phonon coupling strength ($\lambda$) can be estimated using the Debye temperature ($\Theta_D$), which is obtained from the McMillan equation \cite{15}:

\begin{equation}
T_c = \frac{\Theta_D}{1.45} \exp \left( \frac{-1.04(1+\lambda)}{\lambda - \mu^*(1+0.62\lambda)} \right)
\label{eq:macmillan}
\end{equation}

Where $\mu^*$ is the Coulomb pseudopotential, generally taken as 0.13 for transition metals and their compounds \cite{11,12}, and $\lambda$ is the electron-phonon coupling constant. As exhibited in Table \ref{tab:exp}, the magnitude of $\lambda$ is less than one, indicating that the crystals are weak-coupling superconductors. In the optimal composition ZrTe$_{1.85}$Se$_{0.15}$, $\lambda = 0.79$, a value very close to that found in other materials of the same class \cite{12,3e22}.

\begin{table*}
\caption{\justifying Values of the Sommerfeld coefficient, phonon coefficient, Debye temperature, and electron-phonon coupling constant for the single crystal with the highest critical temperature (ZrTe$_{1.85}$Se$_{0.15}$) and for the crystal with the lowest critical temperature (ZrTe$_{1.7}$Se$_{0.3}$). Values for pure ZrTe$_2$ and tellurium-deficient ZrTe$_2$ are also shown.}

\begin{ruledtabular}
\begin{tabular}{cccccc}
\label{tab:exp}
Stoichiometry & $T_c$ (K) & $n(E_F)$ (states/eV) & $\beta$ (mJ/mol K$^4$) & $\Theta_D$ (K) & $\lambda$ \\
\hline
ZrTe$_{1.85}$Se$_{0.15}$ & 4.8 & 0.486 & 1.53 & 156.3 & 0.79 \\
ZrTe$_{1.7}$Se$_{0.3}$ & 2.4 & 0.371 & 0.94 & 201.9 & 0.58 \\
ZrTe$_2$ & without up to 1.8 K & 1.1 & -- & -- & -- \\
ZrTe$_{1.75}$ & 3.2 & 2.1 & -- & -- & -- \\
\end{tabular}
\end{ruledtabular}
\end{table*}

The superconducting dome suggests that Se substitution induces chemical pressure in the crystal lattice. This effect likely modifies both the phonon spectrum and the electronic density of states near the Fermi level. These changes promote the emergence of superconductivity, consistent with BCS theory, where the critical temperature ($T_c$) depends on the Debye frequency ($\omega_D$), the density of states at the Fermi level ($n(E_F)$), and the electron-phonon coupling constant, expressed as
$T_c = \hbar \omega_D \exp(-1/\lambda)$.
However, beyond a certain substitution level, this chemical pressure leads to symmetry breaking that alters the electron-phonon interaction, suppressing superconductivity. Notably, the absence of a specific heat anomaly in this material contradicts the predictions of BCS theory and is compatible with unconventional superconducting behavior.

\begin{figure}
    \includegraphics[width=1.\linewidth]{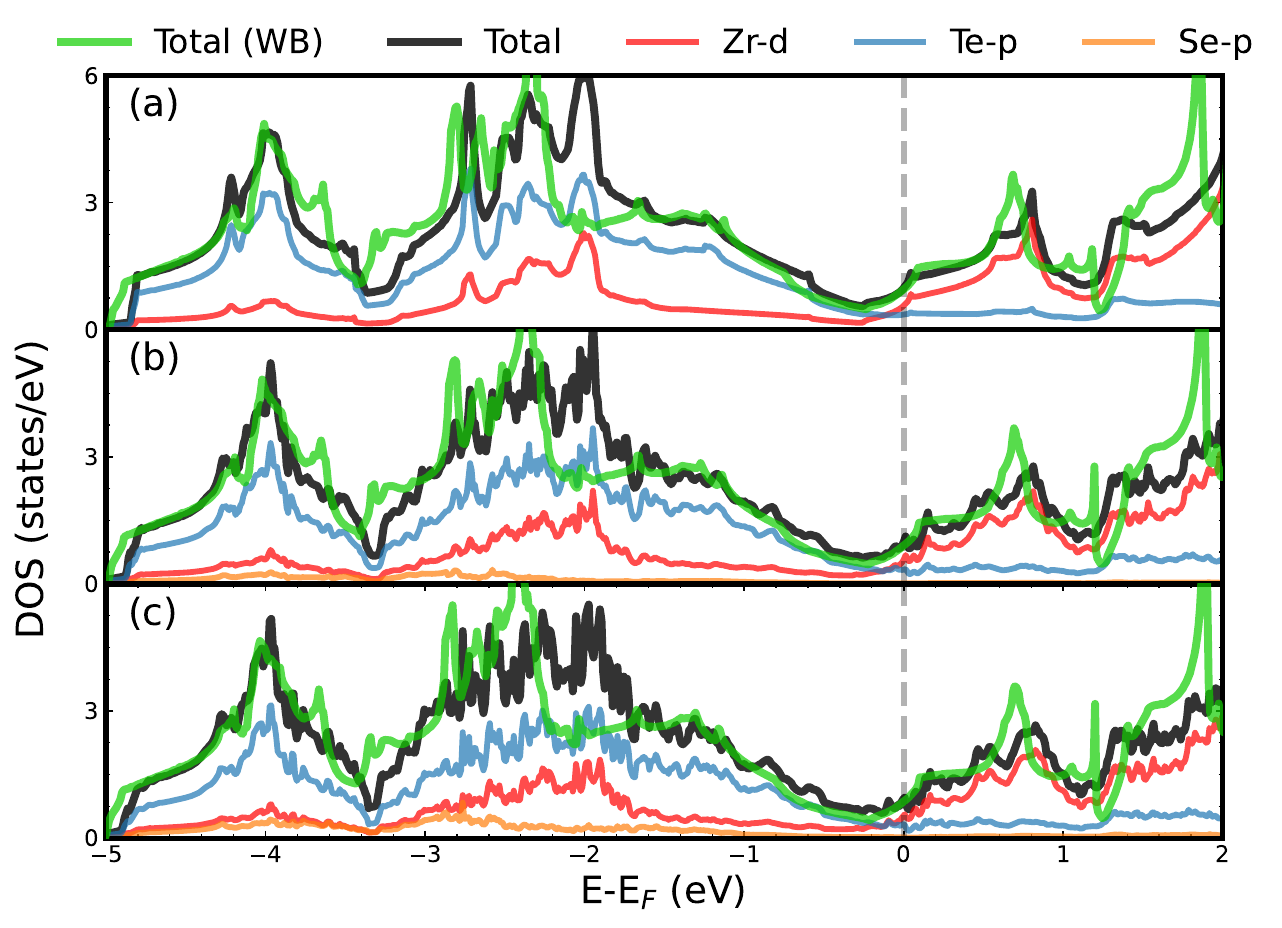}

    \caption{\justifying Evolution of the electronic density of states of ZrTe$_{2-x}$Se$_{x}$ (normalized per unit cell) projected in the orbitals Se/Te-p and Zr-d for (a) $x=0$, (b) $x=0.125$ and (c) $x=0.25$ supercells. In green, we depict the total DOS obtained with the Hamiltonian interpolation.}
    \label{fig:DOS}
\end{figure}

\subsection{First-principles calculations of ZrTe$_{2-x}$Se$_x$}
\subsubsection{Electronic structure evolution}

To gain deeper insights into the electronic structure evolution as a function of composition, we start here by examining its DOS in the range of substitution employed in our experiments. Following Fig. \ref{fig:DOS}, which brings the supercell with SOC taken into account and Wannier interpolation (green curve) results for $x=(0,0.125,0.25)$, it is visible that the change is subtle but noticeable: the valley just below $E_F$ smears out upon substitution, and there is a negligible decrease of the DOS at $E_F$, (from $0.985$ to $0.931\,$eV), slower than the experimental trend but similar to other calculations \cite{PhysRevB.101.165122} that employed GGA-PBE. This discrepancy is explained mostly by the functional employed, but also by our assumption of an ideal crystal structure in calculations. The electronic structure may change with inclusion of disorder scattering such as vacancies and interstitial atoms, that were not considered here. Se-p contribution is mostly perceptible in the valence region, increasing with $x$, inversely proportional to the Te-p contribution, with Zr-d basically unaltered; hybridization between Se-p and other orbitals is visible at $-3\,$eV upwards; and while p orbitals are dominant before $E_F$, Zr-d starts to rise in the valley region, which happens at the same energy in both $x$ levels of Se. This trend is similar to that reported with ZrTe$_2$ under higher pressures \cite{14}.

\begin{figure}
    \includegraphics[width=1.\linewidth]{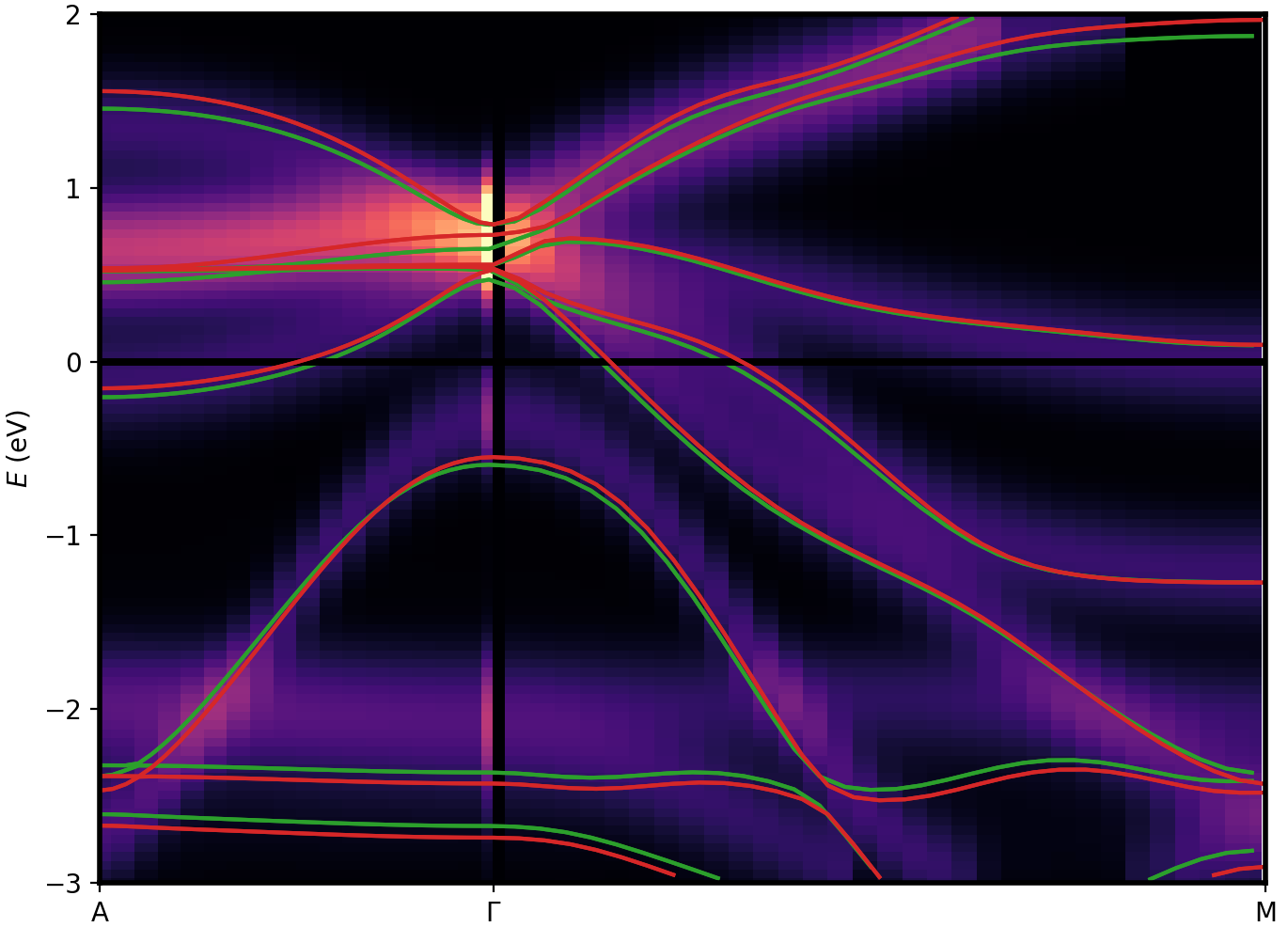}
    \caption{\justifying 
    Electronic structure of ZrTe$_{1.875}$Se$_{0.125}$. The heatmap is  the unfolded, fully-relaxed DFT (QE) supercell calculation, corresponding to the DOS shown in Fig. \ref{fig:DOS}(b). The superimposed green lines give the electronic dispersion using the lattice parameters interpolated from the pristine compounds, while the red lines are the bands using the lattice parameters of the fully-relaxed cell (GPAW+WB). The Fermi energy is set to 0 eV.}
    \label{fig:bands_unf}
\end{figure}

\begin{figure*}[t]
    \includegraphics[width=1.\linewidth]{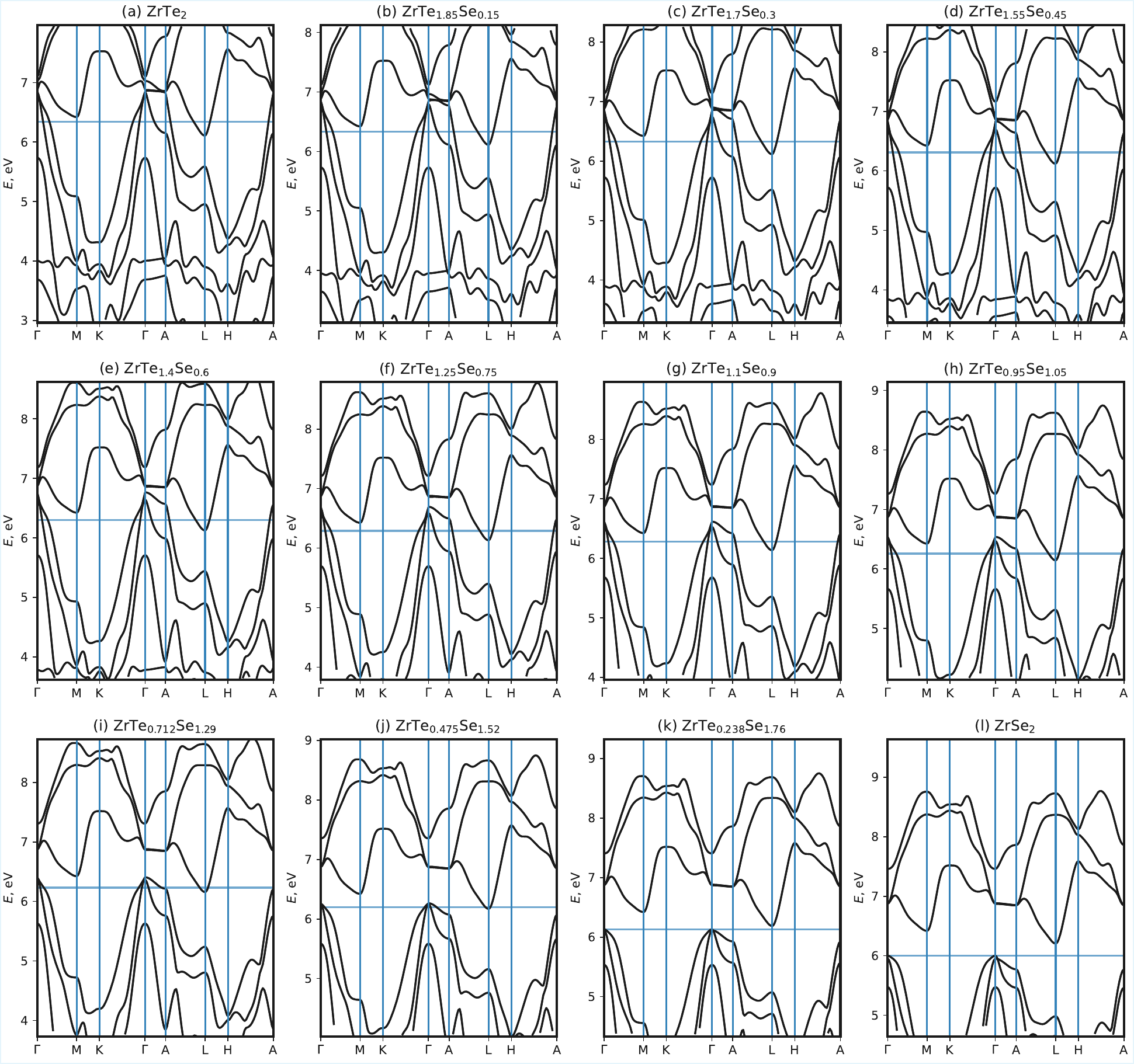}
    \caption{\justifying Interpolated band structures of ZrTe$_{2-x}$Se$_x$ including spin-orbit coupling. The $x$ value update was chosen to be $x+0.15$, in the first and second rows, ranging from (a) $x=0$ to (h) $x=1.05$, and $x+0.2375$ in the last one, from (i) $x=1.29$ to (l) $x=2$. The calculated Fermi energy for each configuration is represented as the horizontal blue line for each plot.}
    \label{fig:bands}
\end{figure*}

In order to obtain the band structures and derived properties for the entire substitution range, we present in Fig. \ref{fig:bands} the result of our interpolation between Wannier Hamiltonians (GPAW+WB) of pristine ZrTe$_2$ and ZrSe$_2$ with effective lattice parameters. For comparison with the usual procedure of recurring to supercell structures and unfolding, we superimpose ZrTe$_{1.875}$Se$_{0.125}$ bands obtained employing the two methods, which are plotted in Fig. \ref{fig:bands_unf}. Overall, the agreement is good in the region near $E_F$ and the first conduction bands, where the Dirac crossing is visible ($A-\Gamma$ path), showing that the topological properties are not affected by this kind of approach. For a better description of the disorder, it would be necessary to assign different $z$-coordinates to the Te and Se atoms, which is impossible within our interpolation method. For further details of interpolation vs supercell differences, the reader can check the Supp. Material \cite{wannierberri-tutorial}.

Returning to the overall picture, as we started our configurations with $z=0.25$, the pristine ZrTe$_2$ electronic structure presented here, in Fig. \ref{fig:bands}(a), agrees with that reported by Tian et al \cite{PhysRevB.102.165149}, with a nodal line semimetal state (NLSM). This enforced degeneracy exists between the highest valence (HVB) and lowest conduction (LCB) bands in $\Gamma-A$ direction, at an energy of $E_F+0.6\,$eV. In this same path, upon Se substitution with a fraction of $x=0.15$ shown in Fig. \ref{fig:bands}(b), the NLSM state is broken, and a type-II Dirac semimetal state (DSM) emerges, with the LCB keeping its original flatband dispersion, but the HVB is shifted downwards in energy after the crossing point in both directions, while no other appreciable effect is observed. This DSM state is preserved when increasing the Se content to $x=0.3$, as the crossing point approaches $\Gamma$ point in Fig. \ref{fig:bands}(c), but at $x=0.45$, the fourfold degeneracy is already gapped out, with a finite energy gap of $50\,$meV between HVB and LCB, that is almost imperceptible in Fig. \ref{fig:bands}(d). After $x=0.45$ up to $x=2$, going towards the pristine ZrSe$_2$,  depicted in Figs. \ref{fig:bands}(e)-(h), we can reinforce with our calculations the already studied transition from semimetal to semiconductor behavior reported by Muhammad et al \cite{10}. It happens while the electron pocket at $L$ shrinks, as the conduction bands are pushed upwards in energy and $E_F$ approaches the top of the hole pocket at $\Gamma$. The unique mismatch here with experiments is that the employed GGA functional severely underestimates the indirect band gap in ZrSe$_2$, being nearly half of the one measured by Mleczko et al \cite{doi:10.1126/sciadv.1700481}, but a more suitable description is provided by other authors using hybrid functionals \cite{https://doi.org/10.1002/pssb.201700033, qin2017strain}. Since the regime above $x=1$ is not the focus of our measurements, the use of a GGA functional is sufficient for comparing the band structures in the experimentally relevant range using Wannier Hamiltonian interpolation.

\begin{figure*}[t]
    \includegraphics[width=1.\linewidth]{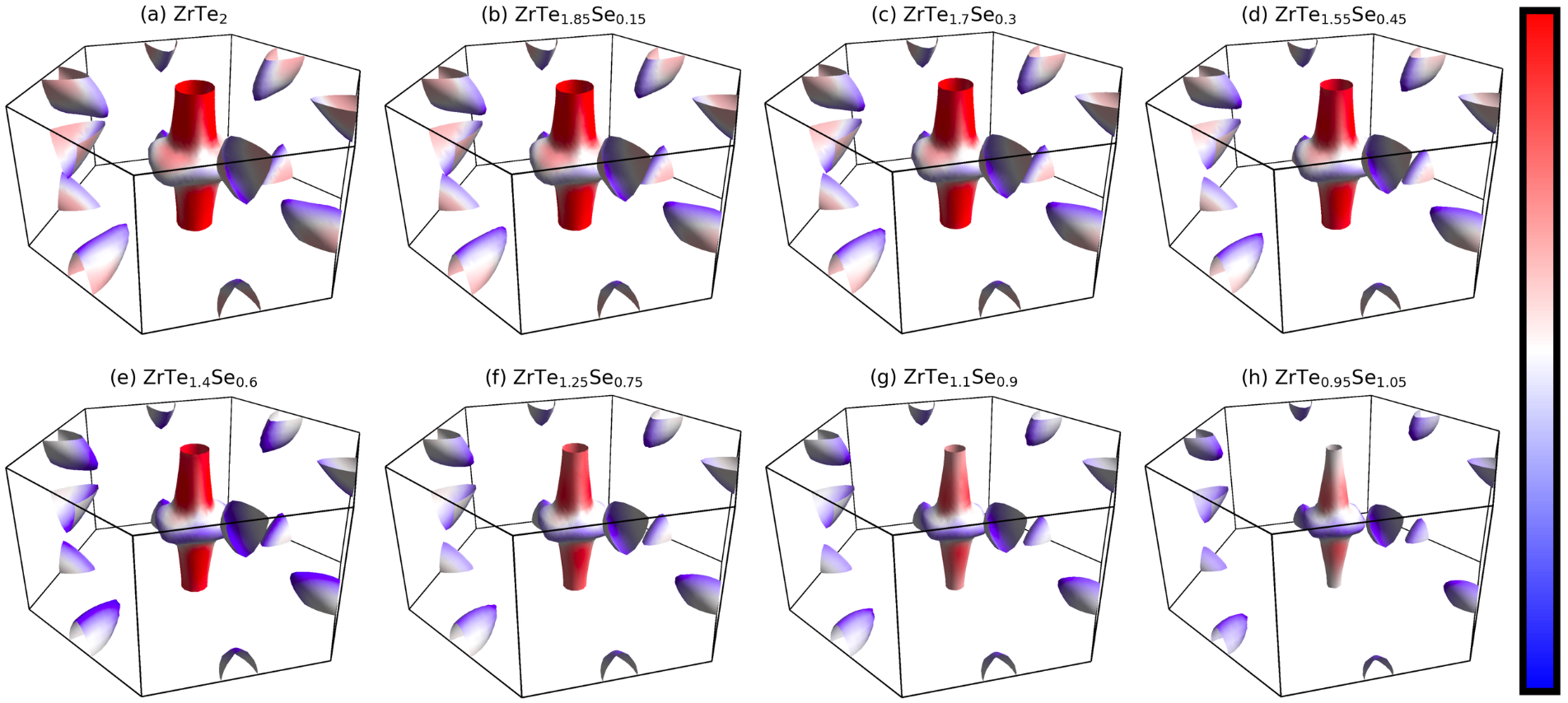}
    \caption{\justifying ZrTe$_{2-x}$Se$_{x}$ Fermi surface evolution through Se substitution. The colorbar indicates the magnitude of the Fermi velocity in arbitrary units.}
    \label{fig:fermi}
\end{figure*}

In this paragraph we direct our discussion to Fig. \ref{fig:fermi}, that shows the Fermi sheets from $x=(0-1.05)$, with their respective Fermi velocities $v_F$. First, the Fermi surface of pristine ZrTe$_2$ consists of three sheets: two hole pockets centered at $\Gamma$, one that has a closed spherical shape, as the band is symmetrical around that point, and the other, due to its band crossing after $A$ point, has its surface as an open tube with a fat center at $\Gamma$, extended in $\mathbf k_z$ direction; and an electron pocket band centered at $L$, giving rise to 12 closed disk-like pockets in the BZ corners. The highest values of $v_F$ are localized in the HVB sheet, when approaching the boundaries at $A$, while the LCB sheet has a low $v_F$ state due to its less dispersive character in $E_F$, as it is near the band concavity. Upon substitution, we verify that the HVB sheet tube remains with a higher $v_F$ up to $x=0.75$, where it suffers a steeper decrease and narrowing in $v_F$, leading to its annihilation at higher $x$ content, where it will be first a closed ellipsoid. Finally, in the LCB sheet disks, $v_F$ starts to vanish within a smaller content of $x=0.45$, where their velocity decreases even more towards zero, but preserving its 3D topology.

\begin{figure*}[t]
    \includegraphics[width=1.\linewidth]{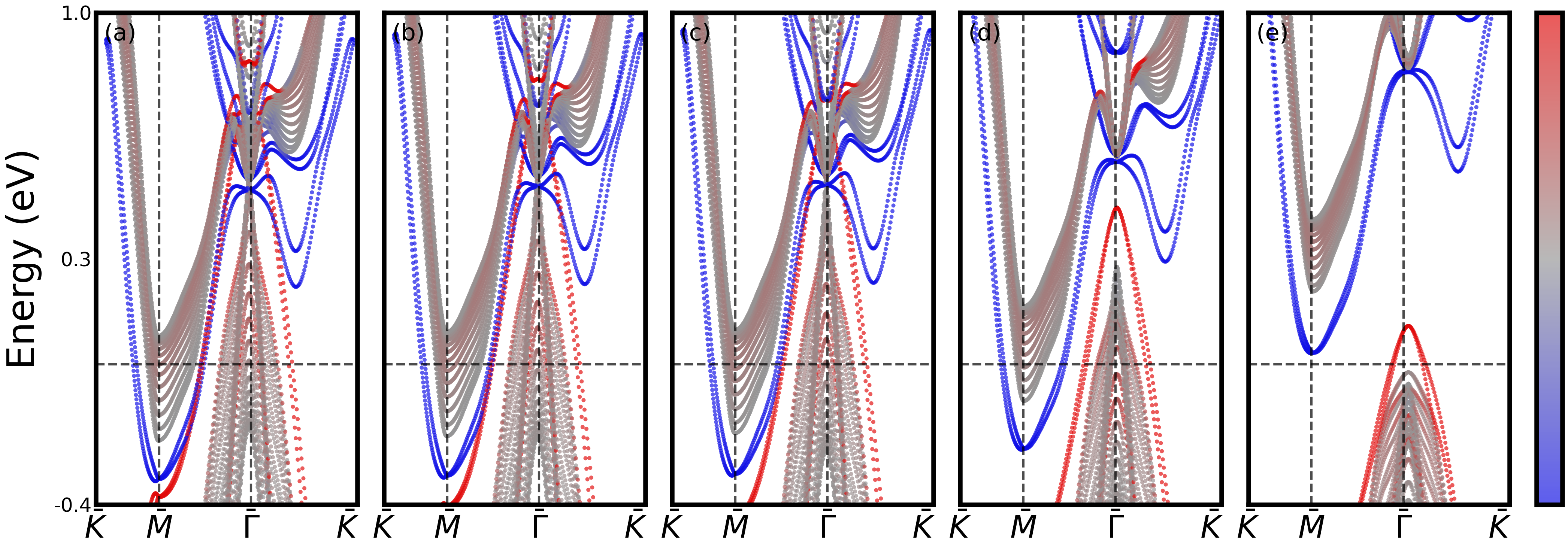}
    \caption{\justifying Calculated surface states on the $(001)$ planes along the high-symmetry path $\bar K\bar M\bar\Gamma\bar K$ for a 14-layer slab, for the configurations: (a) ZrTe$_{2}$, (b) ZrTe$_{1.85}$Se$_{0.15}$, (c) ZrTe$_{1.7}$Se$_{0.3}$, (d) ZrTeSe and (e) ZrSe$_{2}$. Red and blue colors refer to the top (Te) and bottom (Zr) surface weights, respectively. The Fermi energy was set as 0 eV.}
    \label{fig:bands_slab}
\end{figure*}

\begin{figure}
    \includegraphics[width=1.\linewidth]{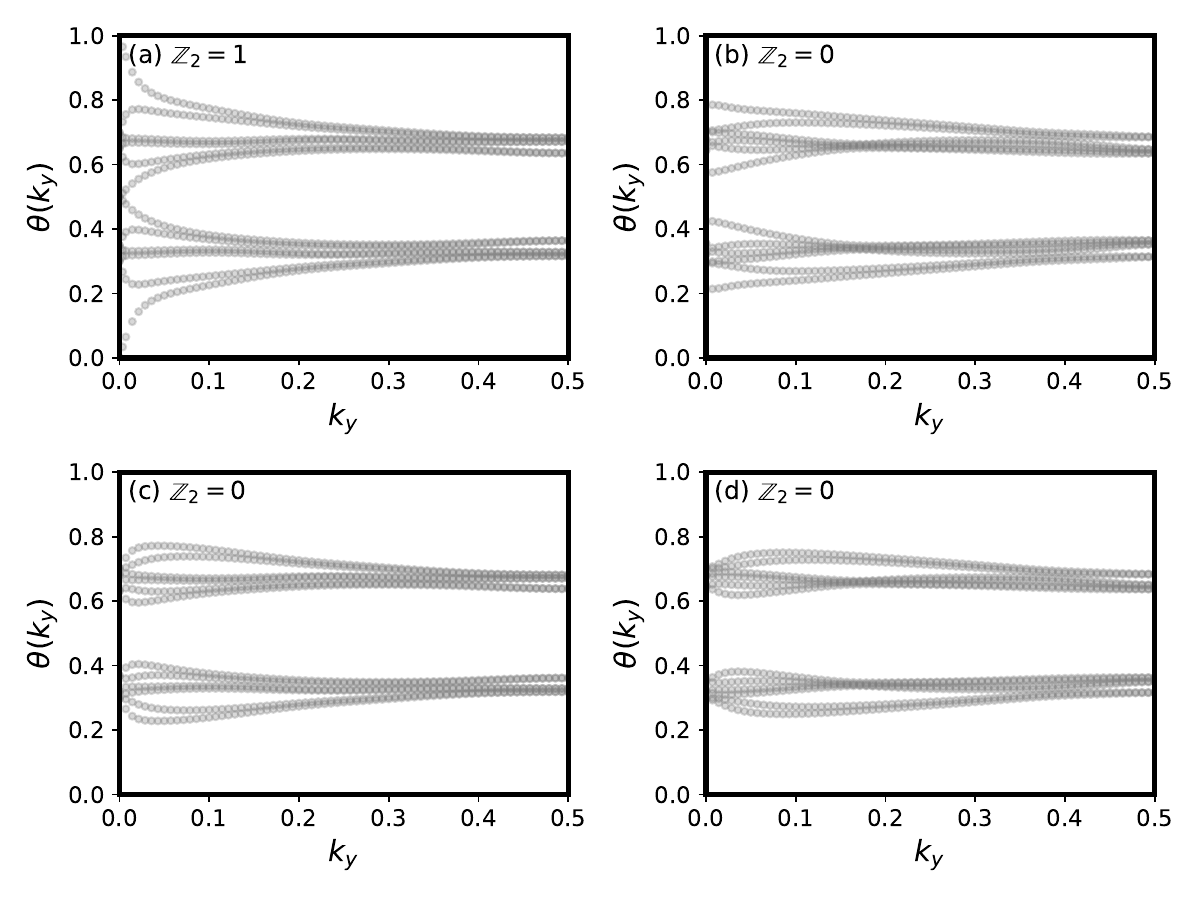}
    \caption{\justifying Non-trivial topological transition depicted by the Wilson loop spectrum of two parallel high-symmetry planes in reciprocal space: (a) and (c): $k_z=0$; (b) and (d):  $k_z=0.5$. Top and bottom rows are ZrTe$_{1.85}$Se$_{0.15}$ and ZrTe$_{1.55}$Se$_{0.45}$, respectively.}
    \label{fig:wilson}
\end{figure}

The surface states (SS), calculated for the plane $(001)$, are shown in Fig. \ref{fig:bands_slab}. A small but converged 14 layer thickness was employed, as more than that will just increase the number of bulk bands at $E_F$, whereas the surface bands remain unchanged. From the plot, we verify that the SS remain in the system up to the complete substitution, with $x=2$. In Figs. \ref{fig:bands_slab}(a)-(c), with $x=(0,0.15,0.3)$ respectively, the nontrivial SS have a wide dispersion, covering an extensive energy range, where they are far above $E_F$ at $\bar\Gamma,\bar K$, but at $\bar M$ the SS are below $E_F$. This feature raises some questions about the possible presence of surface states near $E_F$ in ZrTe$_{0.6}$Se$_{1.4}$, as detected with ARPES in Ref. \cite{10} (because it was supposedly already a semiconductor with this $x$). But, as they are similar to the bulk dispersion, their identification is a more demanding task. At $\bar\Gamma$, the blue SS, corresponding to the bottom surface or Zr termination, occur in a Kramers pair of counterpropagating states with opposite spin in the same energy level as the Dirac crossing point (DP) in the bulk material. This crossing is conserved even after the bulk crossing splitting (DSM to trivial semimetal), consistent with time-reversal symmetry preservation, but it is pushed upwards in energy in the limit $x \to 2$ (Fig. \ref{fig:bands_slab}(e)). It is important to note that at $\bar\Gamma$, the HVB and LCB do not touch each other in any other configuration than pristine ZrTe$_2$ (NLSM config.), and the nontrivial SS occur exactly in the gap between them. The other SS, which are in red denoting the top surface or Te termination, follow the hole pocket centered at $\bar\Gamma$, as they pass through $\bar M$ and surpass the DP in energy. This is consistent with what would be expected from the DOS results, as the conduction bands have Zr-d predominance and the valence bands are Te/Se-p dominated. To gain knowledge in whether this material is a topological or trivial semimetal, as a function of $x$, we depict in Fig. \ref{fig:wilson} the Wilson loop calculation of two TR-invariant planes $(k_z=0.0,0.5)$, that are located above and below the
3D Dirac point of ZrTe$_{2-x}$Se$_x$. This procedure is valid for metallic systems if the bands are gapped on the BZ plane used in the Wilson loop calculation \cite{PhysRevB.95.075146}, and the $\mathbb{Z}_2$ topological invariant is then evaluated for only half of the momentum values (i.e. $0.0-0.5$). Thus, for $x\leq0.45$, we verify that the system has a nontrivial $\mathbb{Z}_2=1$ in plane $k_z=0.0$ and a trivial one at $k_z=0.5$ (represented by ZrTe$_{1.85}$Se$_{0.15}$ in Figs. \ref{fig:wilson}(a)-(b)), which is the same behavior of other materials hosting protected time-reversed partners such as BiNa$_3$ \cite{na3bi} and Cd$_3$As$_2$ \cite{Liu2014Cd3As2}, despite the fact that their Dirac points are very near $E_F$. This evaluation can support the claim of a persistent DP up to higher concentrations, with nontrivial topological invariants $(1;000)$. The values of $(v_0;v_1v_2v_3)$ are in agreement with those reported by \cite{14}, at $0$ GPa, but here the same invariants are maintained with the increasing chemical pressure, whereas with external pressure the system has multiple phase transitions. After that value of $x$, all the evaluations resulted in trivial indicators up to $x=2$ $(0;000)$, shown in Figs. \ref{fig:wilson}(c)-(d); they can be viewed in more detail in the Supp. Material Fig. S2. With these results, we can state that there is a coexistence of superconductivity and nontrivial band topology, which could be delimited by the phase transition from enforced semimetal to trivial semiconductor, even though we don't have a sufficient degree of accuracy to  determine the exact $x$-value in which this transition occurs.

\subsubsection{Implications for the superconductivity}

\begin{figure}
    \centering
    \includegraphics[width=1.\linewidth]{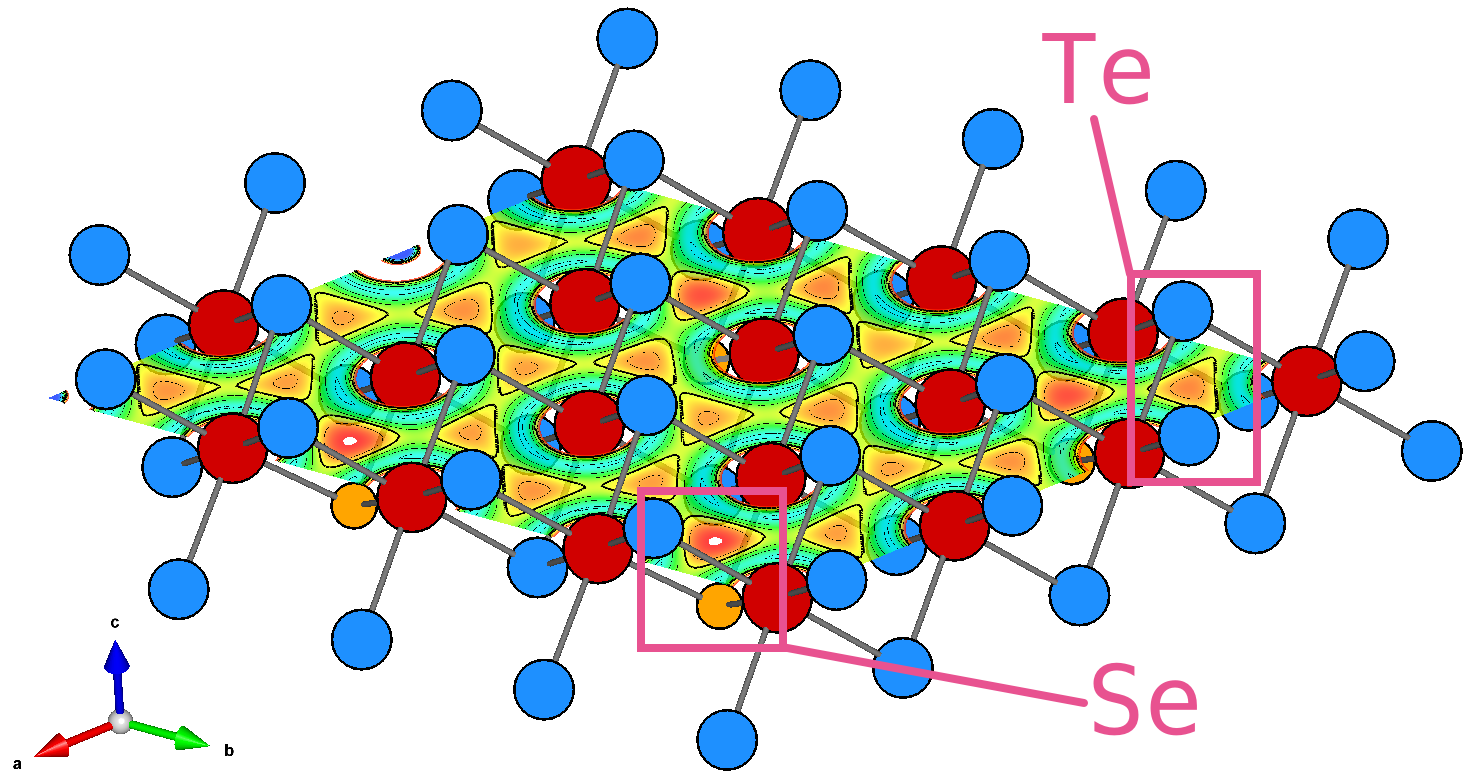}
    \caption{\justifying The calculated electronic
localization function (ELF) of ZrTe$_{1.875}$Se$_{0.125}$ for the plane $(001)$ (Zr plane). The ELF range was set from $0$ to $0.7$.}
    \label{fig:elf_zrte2sex}
\end{figure}

By inspecting the structural effects upon increasing $x$, from our relaxed structures, it is worth noting that the main modification caused by Se substitution is that of changing Te-Te out-of-plane distance $d_{\text{Te-Te}}$, which is further reduced for $x=0.125$ and $0.25$. To elucidate this, the pristine ZrTe$_2$ has all $d_{\text{Te-Te}}=3.798\,$\AA, while the ZrTe$_{1.875}$Se$_{0.125}$ has $d_{\text{Te-Te}}=3.76\,$\AA{} between a layer where Se joined the crystal, and $d_{\text{Te-Te}}=3.790\,$\AA{} in a layer without Se. This bond length contracts to $d_{\text{Te-Te}}=3.748\,$\AA{} in ZrTe$_{1.75}$Se$_{0.25}$, and could increase Te chain interactions. As the highest two valence bands are mostly Te-p dominant and strongly dependent of this level of hybridization, mainly between Te-Zr (that follows the same pattern with $x$), increasing the inter-orbital coupling ultimately will lift the HVB and LCB degeneracy, turning it into the trivial semimetal phase. A similar behavior occurs in ZrTe$_{3-x}$Se$_x$, in which a BCS superconductivity with intraband coupling was reported, but the optimal range is much more restrictive ($x\leq0.1$), as the DOS decreasing at $E_F$ weakens the SC phase \cite{9,PhysRevLett.122.017601,cpb_zhang_zrte3}. In respect to the competition between CDW and SC orders, we touch upon this question with the idea that a small $x$ is sufficient to cause an electronic disorder scattering in the system such that it suppresses the hidden CDW \cite{Li2017TaSe2Sx,vv94-wx44,lz1c-bmpw}. To elucidate this, we calculated the electron localization function (ELF) \cite{elf-paper,elf1} for the 2D Zr plane of ZrTe$_{1.875}$Se$_{0.125}$ supercell, depicted in Fig. \ref{fig:elf_zrte2sex}. The influence of the higher electronegativity from Se is visible, hosting a much more localized electronic cloud that effectively breaks the underlying periodicity. What motivates this logic is a work for $s$-wave superconductors, in which Banerjee et al \cite{PhysRevB.98.104206} proposed, using an extended Hubbard model with a disorder potential ($V$). This could be a general trend, where charge modulations loose their phase coherence with increasing $V$ (in our case, it is tuned by the $V(x)$ of Se-p), and beyond a certain threshold they will not interfere in the SC properties, even if coexisting in a local short-range. Also, via ELF, Liu et al showed that HfTe$_3$ filamentary SC was achieved, supressing CDW that have its origin in the induced localization between Te atoms from different site positions under pressure \cite{Liu2021HfTe3}, stabilizing the SC state.

In light of the experimental results, we can argue that the superconducting state could be similar to the situation reported by Hossain et al \cite{Hossain2025Superconductivity} in the bulk topological semimetal ZrAs$_2$, which hosts a superconductivity confined to its surface states within the $ab$ plane. Another argument is raised by comparing the SC of ZrTe$_{1.85}$Se$_{0.15}$ to that of Cu$_x$ZrTe$_{2-y}$ \cite{PhysRevB.95.144505}. Our material shows a weak diamagnetic response in the order of $10^{-3}$ smaller than that of the intercalated one and is in the weak electron-phonon coupling regime ($\lambda<1$). Also, as the increasing $x$ content of ZrTe$_{2-x}$Se$_x$ has the effect of suppressing $N(E_F)$ in the bulk, because the HVB and LCB move apart in energy, which is detrimental for the SC ordering, the $(001)$ surface states remain persistent for a broader range, specially those in Zr termination. That is exactly the plane in which CDW ordering occurs. The emergence of SC could also be related to its Fermi surface quantum geometry \cite{quantum-geometry1}, as our material has nontrivial band topology, which was shown to improve the electron-phonon coupling even in materials with dispersive bands, for example MgB$_2$ \cite{quantum-geometry2}.

As possible future directions, we have in a first case ZrTe$_{2-x}$Se$_x$ being a platform for hosting topological SC, considering the simple fact that it has nontrivial SS at $E_F$ and is an intrinsic superconductor, which places it as a potential material for the study of supersymmetry and Majorana bound states \cite{PhysRevLett.109.237009}. Despite the fact that it is not a stoichiometric superconductor, such as noncentrosymmetric PbTaSe$_2$ \cite{PhysRevB.93.245130}, and centrosymmetric TaC \cite{PhysRevB.101.214518} and $\beta$-PdBi$_2$ \cite{Sakano2015,Powell2025}, its SS are separated from bulk states in $\mathbf k$ space and they have a fully gapped SC (results discussed below in Fig. \ref{fig:bdg}(a)), which are necessary for helical surface states \cite{Xu2014,PhysRevLett.100.096407}. For example, the spin texture of the 2 surface bands of the Zr termination at $E_F$, shown in Fig. \ref{fig:bdg}(b), exhibit the 6 pockets at $\bar M$ that generates the spin-locked surface Dirac cone at $\bar\Gamma$ far from $E_F$, as they are very close together in energy, but remain distinguishable. The lack of experimental confirmation from ARPES measurements will motivate a future work. 

Now we discuss the second possibility, that of engineering heterostructures combining different materials \cite{Lutchyn2018Majorana,Kezilebieke2020TopologicalSC,PhysRevLett.114.017001,doi:10.1126/science.1222360}, where we will use something similar to NbSe$_2$/CrBr$_3$ here. Also, very recently, Islam et al \cite{islam2026emergentsuperconductivitynonreciprocaltransport} found a SC state with magneto-chiral anisotropy in the transition region of ZrTe$_2$/CrTe$_2$ ($T_c\approx10\,$K), a feature observed in topological insulators. Thus, we approach this with the method of inducing a s-wave pairing in an heterostructure with a ferromagnetic insulator material to induce surface Zeeman spin splitting, in order to simulate the SC ground state \cite{Hu2025}. Here we start assuming a conventional BCS-like Cooper pairing, based on the experimental input. Within this method, we construct a 2D slab system $H^{slab}_{mn}(\mathbf k_{||})$, with matrix elements $H^{ij}_{mn}$, for the composition with highest $T_c$, ZrTe$_{1.85}$Se$_{0.15}$. Starting from its $H_{mn}(\mathbf R)$, and applying open boundary conditions for $\mathbf R^{'}_3$, we get:
\begin{align}
H^{ij}_{mn}(\mathbf k_{||})=\sum_{\mathbf R^{'}}H_{mn}(\mathbf R^{'})e^{-i\mathbf k_{||}\cdot\mathbf R^{'}},
\end{align}
where $\mathbf R^{'}=\{\mathbf R^{'}_1,\mathbf R^{'}_2,(i-j)\mathbf R^{'}_3\}$ and $i,j$ are the indices of the 14 layer slab. Superconducting pairiting is then induced across the entire system, and the Bogoliubov-de Gennes (BdG) Hamiltonian \cite{Sato_2017}, can be derived as follows, using the Nambu basis:
\begin{align}
H^{slab}_{BdG}=
\begin{pmatrix}
H^{slab}_{mn}(\mathbf k_{||},\mathbf M)-\mu & \Delta \\
\Delta^\dagger & -H^{slab}_{mn}(\mathbf k_{||},\mathbf M)^*+\mu
\end{pmatrix}.
\end{align}
The $H^{slab}_{BdG}$ spectrum is calculated for a surface spin splitting $M_z=0$, shown in Fig. \ref{fig:bdg}(a), with an intra-orbital s-wave pairing $\Delta=0.001\,$eV and a chemical potential $\mu=0\,$eV, with WT default parameters for decay (corresponding to values for Bi$_2$Se$_3$). In general, this is to show that, if the superconducting phase is in the bulk, the induced SC in the SS by the proximity effect will be predominant as the bulk states are gradually spaced from $E_F$. Now, by constructing a topological phase diagram with both $M_z$ and $\mu$ varying, shown in Fig. \ref{fig:bdg}(c), we verify that topological superconducting states may be achieved within a relatively small splitting. As our compound is non-stoichiometric, the $\mu$ parameter can be tuned by the substitution $x$ fraction, as long as it stays in the nontrivial DSM phase.

\begin{figure*}
    \includegraphics[width=1.\linewidth]{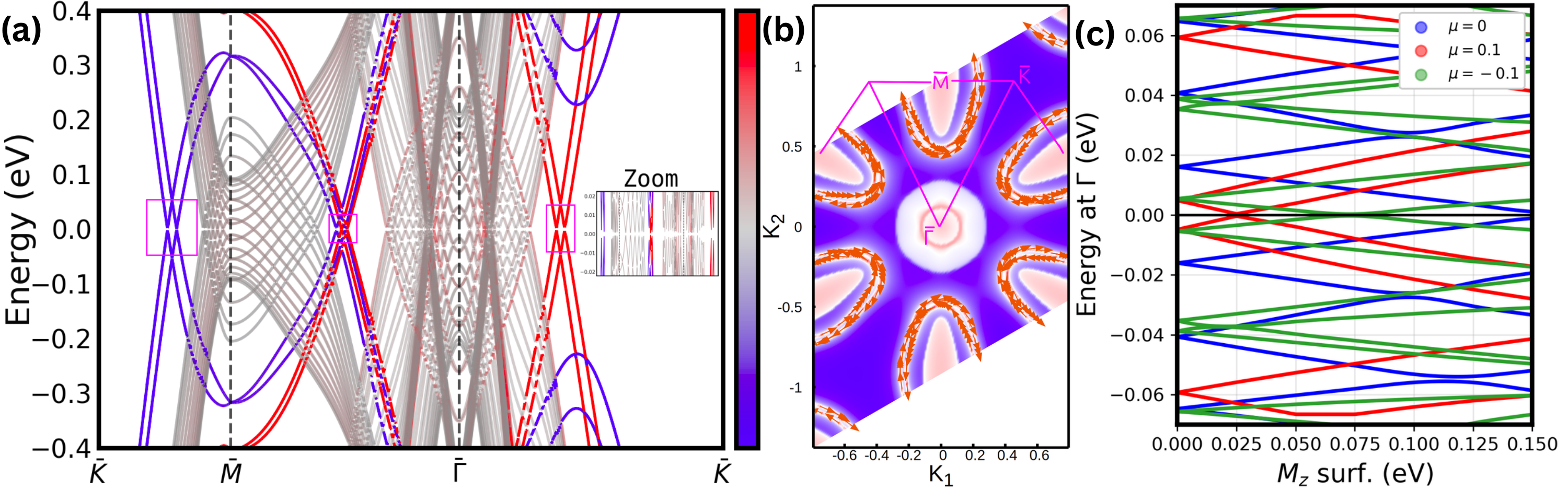}
    \caption{\justifying(a) Bogoliubov-de Gennes (BdG)
Hamiltonian spectrum for the 14 slab of ZrTe$_{1.85}$Se$_{0.15}$. At zero eV we have the superconducting gap $2\Delta=2.0\,$meV, and an inset zoom was applied to the nodeless gap. (b) The spin texture of the surface states of $(001)$ surface from Zr-termination, at $E=E_F$. (c) Energy of the BdG Hamiltonian of the system at
$\Gamma$-point as a function of the surface spin splitting $M_z$, for 3 different values of the chemical potential $\mu=(-0.1,0,0.1)\,$eV.}
    \label{fig:bdg}
\end{figure*}

\section{Summary}

In summary, this work demonstrates that the substitution of tellurium with selenium in ZrTe$_{2-x}$Se$_x$ induces an internal chemical pressure that systematically modulates the material's electronic ($\gamma$) and phononic ($\beta$) contributions. Our results reveal the emergence of superconductivity with a $T_c$ that is optimized at an intermediate Se concentration, suggesting the existence of a superconducting dome as a function of selenium substitution. However, a central observation is the absence of two common signatures of conventional superconductors: the specific heat anomaly and a sharp diamagnetic response. The lack of these features, which are hallmarks of conventional BCS-type superconductors, raises profound questions about the nature of the superconducting state in this system. Therefore, with extensive ab initio calculations and Wannier interpolation, we found that the nontrivial DSM coexists exactly with the SC phase, and after a value of $x=0.4$ the system becomes a trivial semimetal. Thus, its topological electronic structure, with $\mathbb Z_2=1$, put this non-stoichiometric material as an interesting playground for topological superconductivity, or even may be the case of being an intrinsic topological superconductor. 

\section*{Data availability}
All the relevant computational data of this research is provided in this GitHub repository: \href{https://github.com/cauaschuch/paper-ZrTe_2-xSe_x}{\texttt{paper-ZrTe2-xSex}}, which after the paper acceptance will be open sourced at Zenodo. Experimental data will be made available upon reasonable request.

\begin{acknowledgments}
The research was carried out using high-performance computing resources made available by the Superintendência de Tecnologia da Informação (STI), Universidade de São Paulo. We acknowledge the support of the INCT project Advanced Quantum Materials, involving the Brazilian agencies CNPq (Proc. 408766/2024-7), FAPESP (Proc. 2025/27091-3), and CAPES. L.T.F.E acknowledges CPNq Grant 311756/2022-0.

\end{acknowledgments}

\bibliography{apssamp}

\end{document}